# Towards Biosensing Strategies Based on Biochemical Logic Systems

Evgeny Katz
*Department of Chemistry and Biomolecular Science, Clarkson University, Potsdam, NY 13699, USA*
ekatz@clarkson.edu

Vladimir Privman
*Department of Physics, Clarkson University, Potsdam, NY 13699, USA*
privman@clarkson.edu

Joseph Wang
*Department of Nanoengineering, University of California at San Diego, La Jolla, CA 92093, USA*
josephwang@ucsd.edu

## Abstract

Recent advances in biochemical computing, i.e., information processing with cascades of primarily enzymatic reactions realizing computing gates, such as **AND**, **OR**, etc., as well as progress in networking these gates and coupling of the resulting systems to smart/responsive electrodes for output readout, have opened new biosensing opportunities. Here we survey existing enabling research results, as well as ideas and research avenues for future development of a new paradigm of digitally operating biosensors logically processing multiple biochemical signals through Boolean logic networks composed of biomolecular reactions, yielding the final output signals as YES/NO responses. Such systems can lead to high-fidelity biosensing compared to common single or parallel sensing devices.

## 1. Introduction

We outline the conceptual foundations of the novel approach to biosensing based on multi-step processing of biochemical signals through biocatalytic/biorecognition processes, adapting ideas recently developed in the field of biocomputing (biomolecular logic). Signal processing performed by biochemical means can be followed by transduction of the output chemical signals to the final electronic signals by electrochemical methods. Alternatively, the chemical output signals can be directly used for activation of "smart" chemical actuators based on stimuli-responsive materials. Thus, biosensors will digitally process multiple biochemical signals according to built-in Boolean logic "programmed" in the biomolecular composition of the systems. The information processing will be performed by biochemical means based on the biocomputing approach without electronic components. A byproduct will be having less "wires" and "batteries" (reducing the overall need for electrical power supplies) at those stages of information processing that are carried out "on-site" in implantable devices, and other applications.

The digitally processed information will produce the final output in the YES/NO form, allowing direct coupling of the signal processing with chemical actuators resulting in integrated biosensor-bioactuator ("Sense/Act/Treat") systems. The information processing systems based on biocatalytic reactions performed by enzymes or bioaffinity processes upon antigen-antibody coupling will mimic computing networks where input/output signals will be represented by biochemical means.

This approach to digital biocomputing integrates biological and electronic/computing concepts. Information processing is accomplished in life sciences and biotechnology in two different ways. On one hand, living organisms posses various functions that "process information" on many levels, from transcription and signaling pathways to interactions with the external world. Nature uses information processing paradigms different from those in electronic computers. On the other hand, processing of information in man-made sensor/actuator systems coupled with ordinary electronics involves increasingly complex "decision making" and interfacing in signal processing and sensing, benefiting from the paradigm of digital computing and Boolean logic implemented in the electronic parts of the system. In the present approach we combine the specificity of biomolecules with digital information processing, benefiting from both. Indeed, scaling up the complexity of the biomolecular systems from individual logic gates to their networks is a particularly challenging goal in this field of research.

## 2. Enzyme-based logic gates and networks

Extensive research in unconventional chemical computing resulted in the formulation of various logic systems processing chemical signals according to Boolean logic using chemical means [1-4]. The use of biomolecular/biological tools for logic operations [5-7], such as proteins [7-12], DNA [13-15], RNA [16] and even whole cells [17], broadened the research area of information processing benefiting from the specificity of biomolecular interactions. Recently realized enzymatic logic gates allowed information processing mimicking various Boolean logic operations: **AND**, **OR**, **XOR**, **NAND**, **NOR**, etc. [18,19], see Fig. 1. Scaling up the complexity of the enzyme logic systems by concatenating the logic gates resulted in networks performing simple arithmetic operations [20] and multi-signal processing [21,22], see Fig. 2, through biocatalytic reactions. Simultaneous processing of multiple biochemical signals applied in different combinations allowed logic processing of the signal patterns according to the built-in "program" [21,22], see Fig. 3.

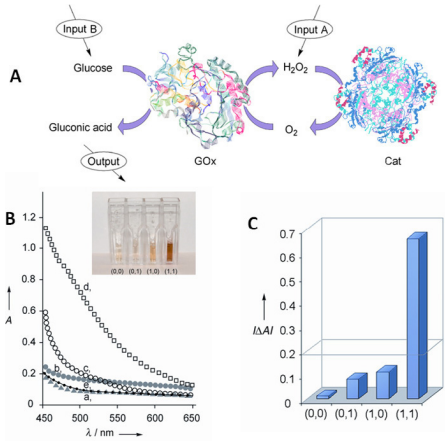

**Fig. 1.** (A) Enzyme based **AND** logic gate. (B) Optical changes in the system generated in the presence of different combinations of the input signals: a) **0,0**; b) **0,1**; c) **1,0**; d) **1,1**; e) prior to the signals application. Inset: Solutions featuring absorbance changes in the system. (C) Bar diagram showing the absorbance changes as a function of the input signals. Adapted from [19].

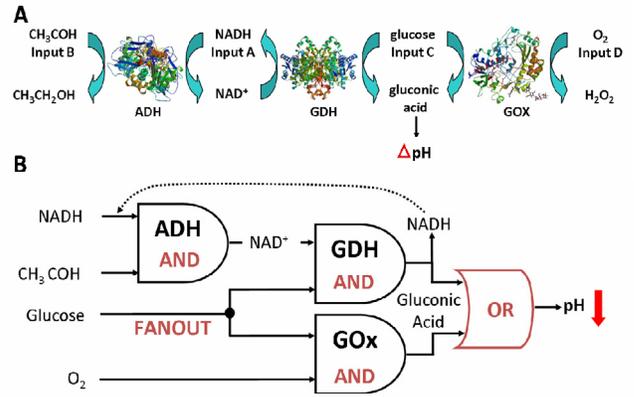

**Fig. 2.** (A) The multi-gate / multi-signal processing enzyme logic system producing *in situ* pH changes as the output signal. (B) The equivalent logic circuitry for the biocatalytic cascade. Adapted from [21].

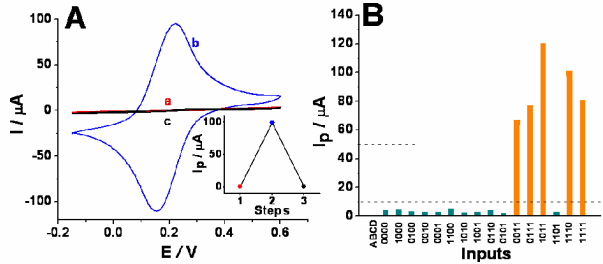

**Fig. 3.** (A) Cyclic voltammograms obtained for a signal-responsive switchable electrode controlled by the enzyme logic network: a) initial OFF state, b) ON state enabled by the input combinations, c) *in situ* reset to the OFF state. Inset: reversible current changes upon switching the electrode ON-OFF. (B) Anodic peak currents, $I_p$, for the 16 possible input combinations. Adapted from [21].

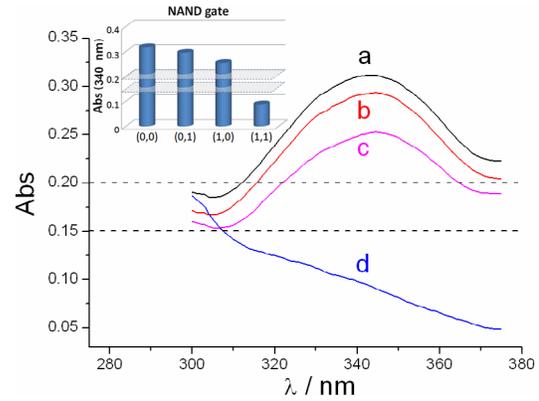

**Fig. 4.** Optical responses generated by the **NAND** logic gate presented in Fig. 5, for input signals of maltose and phosphate: a) **0,0**; b) **0,1**; c) **1,0** and d) **1,1**. Adapted from [24].

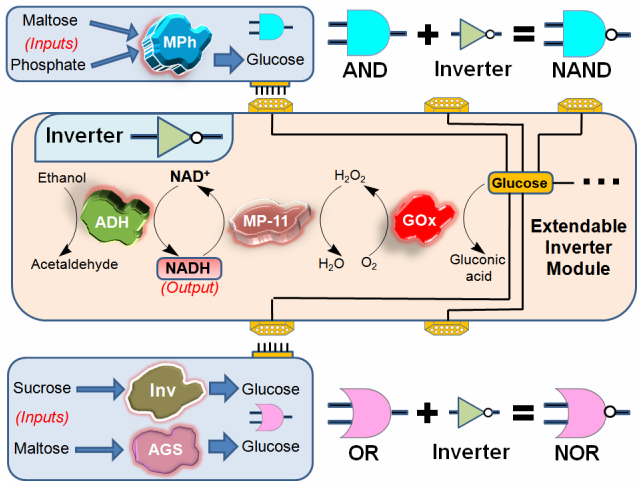

**Fig. 5.** Modular design of the enzyme-based **NAND** and **NOR** gates. Adapted from [24].

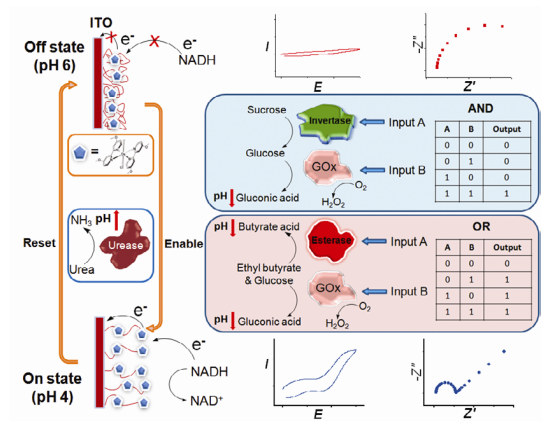

**Fig. 6.** Logic operations **AND/OR** performed by the enzyme-based systems resulting in the **ON** and **OFF** states of the bioelectrocatalytic interface followed by the **Reset** function to complete the reversible cycle. Adapted from [33].

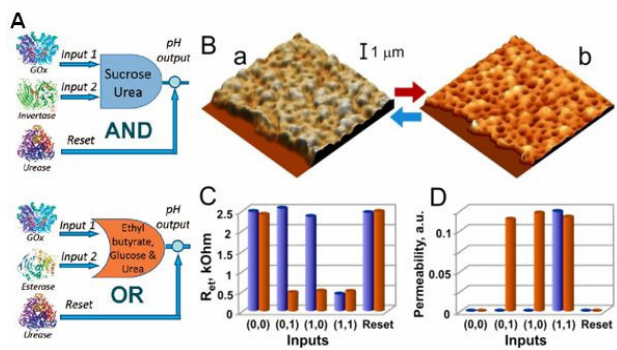

**Fig. 7.** (A) Enzyme logic gate operations followed by the enzyme reset function coupled with signal-responsive materials. (B) A polymer membrane switched between closed (a) open (b) states upon getting pH signals from the enzyme logic gates. (C) Impedance values switched by the membrane deposited on an electrode. (D) Permeability of the membrane switched by the logic signals. Adapted from [28].

Complex functions of biocomputing devices based on logic networks were achieved, e.g., a biomolecular keypad lock performing logic **Implication** function was developed [23]. Following the concepts of electronic computing systems, the first steps have been taken to develop the biochemical versions of integrated computing circuitries [24]. Universal logic gates **NAND**/**NOR** were developed, see Fig. 4, and the possibility of a modular design of biocomputing networks composed of standard repeatable units was demonstrated, see Fig. 5. The use of these universal logic gates will allow future designs of different logic schemes based on the same standard units.

The systems for amplification of biochemical signals and their transformation from one reactant to another were explored in order to design integrated biochemical circuits with standard repeating modules fulfilling Boolean requirements for networks [24]. This research started as a fundamental study of novel concepts in unconventional computing but has already found practical applications in biosensing [25,26].

## 3. Coupling of enzyme gates and networks with signal sensing and actuating interfaces

Signal processing by the enzyme-logic gates and their networks was used to switch signal-responsive materials between different states, thus allowing transduction of chemical output signals into changes of material properties [27-31]. The coupling between signal-processing and stimuli-switchable systems was achieved by pH changes generated *in situ*. Logic **AND**/**OR** operations performed by the enzyme systems were coupled with the reversible transition of polymer-modified electrodes to activate/deactivate bioelectrocatalytic reactions at sensing interfaces (e.g., glucose [32] or NADH [33] oxidation). The pH changes governing the interfacial properties were produced *in situ* in bulk solutions [21,32-34] or locally directly at the modified surface [35]. The developed approach resulted in the switchable sensing interfaces [33], see Fig. 6, and chemical actuators, e.g., switchable membranes [28], see Fig. 7, activated on demand by signals received from the enzyme logic systems.

## 4. Optimization of signal processing and noise reduction in enzyme logic systems

The increasing complexity of enzyme-based logic networks presents interesting theoretical modeling challenges. These include: analog noise suppression through gate and network optimization; control of digital noise through network redundancy; and development of biochemical filters, rectifies and other Boolean and non-Boolean logic gates and network elements. Within the analog/digital information processing paradigm, error buildup can be suppressed by gate optimization for reduction and control of the analog noise amplification [36], as well as by network design and/or network topology [37]. For larger networks, another, "digital" [36-38] mechanism of noise amplification emerges which can be controlled by utilizing redundancy in network design and requires truly digital information processing with appropriate network elements for filtering, rectification, etc. Biochemical computing gates and networks realized and theoretically modeled in our work, have allowed exploration of aspects of design and optimization issues related to suppression of the "analog" noise amplification [36-41] by modeling the biochemical reactions involved within the rate-equation approach and adjusting the "gate machinery" to achieve the optimal shape of the response-surface function, $z = F(x, y)$, where $x, y, z$ are the two inputs and the output variables scaled to the "logic" range between Boolean **0** and **1**, e.g., Figs. 8-9 (additional details can be found in [36-41] and some are outlined below in Sect. 7). The goal is to practically avoid analog noise amplification as measured by the fractional growth of the width of the noise distribution; see Fig. 9.

Recently, an enzymatic system with a self-promoter substrate (typical of allosteric enzymes) was also explored [40]: Electrode-immobilized enzyme glucose-6-phosphate dehydrogenase (G6PDH) was used to catalyze an enzymatic reaction which carries out an **AND** logic gate. A kinetic model was developed and utilized to evaluate the extent to which the experimentally realized gate is close to optimal, which in this case corresponds to virtually no analog noise amplification.

We have found [36-41] that optimization of gates one at a time is not always a straightforward procedure since the gate response function depends parametrically on several biochemical quantities some of which are difficult to change in a broad range of values. Therefore, we have also explored [37,41] optimization of a *network* of enzymatic reactions as a whole: We seek a few-parameter *modular* description of the network elements that will allow us to "tweak" the relative gate activities in the network to improve its stability. We used this approach for the network of three **AND** gates, shown in Fig. 10 (details are given in [37,41]).

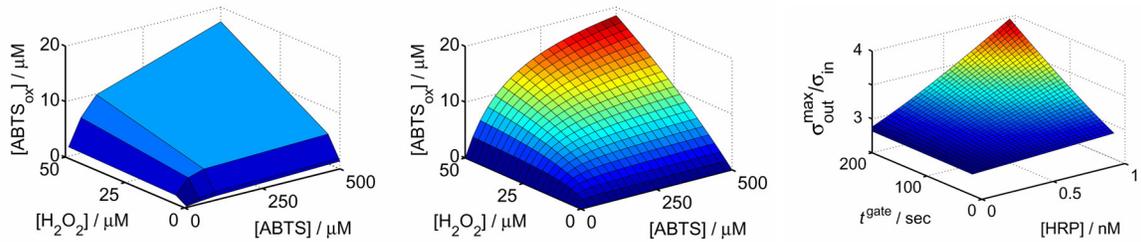

**Fig. 8.** Reaction catalyzed by the enzyme horseradish peroxidase (HRP), with hydrogen peroxide as the substrate and 2,2'-azino-*bis*(3-ethylbenzthiazoline-6-sulphonic acid) (ABTS) as the co-substrate. Measured (left) and numerically fitted (center) response surface (here with variables not scaled to the "logic" range between **0** and **1**) for this **AND** gate. Right: The gate function quality measure, $\sigma_{out}^{max}/\sigma_{in}$, as a function of the enzyme concentration and reaction time. Here $\sigma_{in}, \sigma_{out}^{max}$ are the statistical spreads of the input and output signals. Experiment and model details can be found in [39,41]: ABTS was found to yield large analog noise amplification.

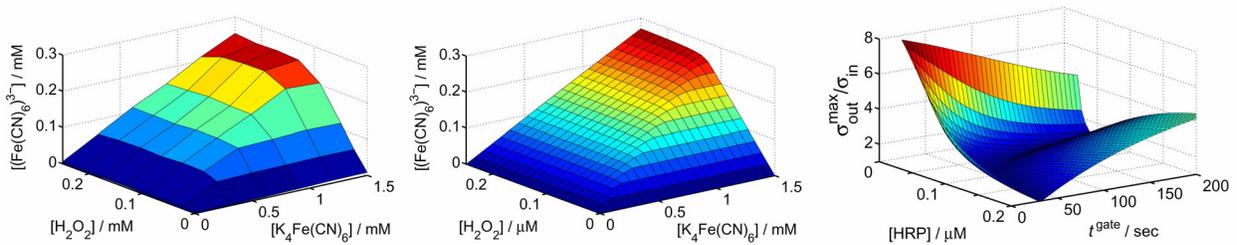

**Fig. 9.** Similar to Fig. 8, measured (left panel) and numerically fitted (center panel) response surface for the HRP-based logic gate, but now with co-substrate $K_4Fe(CN)_6$ (ferrocyanide) as one of the inputs. Right panel: Surface plot of the gate function quality measure, $\sigma_{out}^{max}/\sigma_{in}$, as a function of the enzyme concentration and reaction time. This optimized gate allows parameter selection to have practically no noise amplification ($\sigma_{out}^{max}/\sigma_{in}$ close to 1); see [39,41].

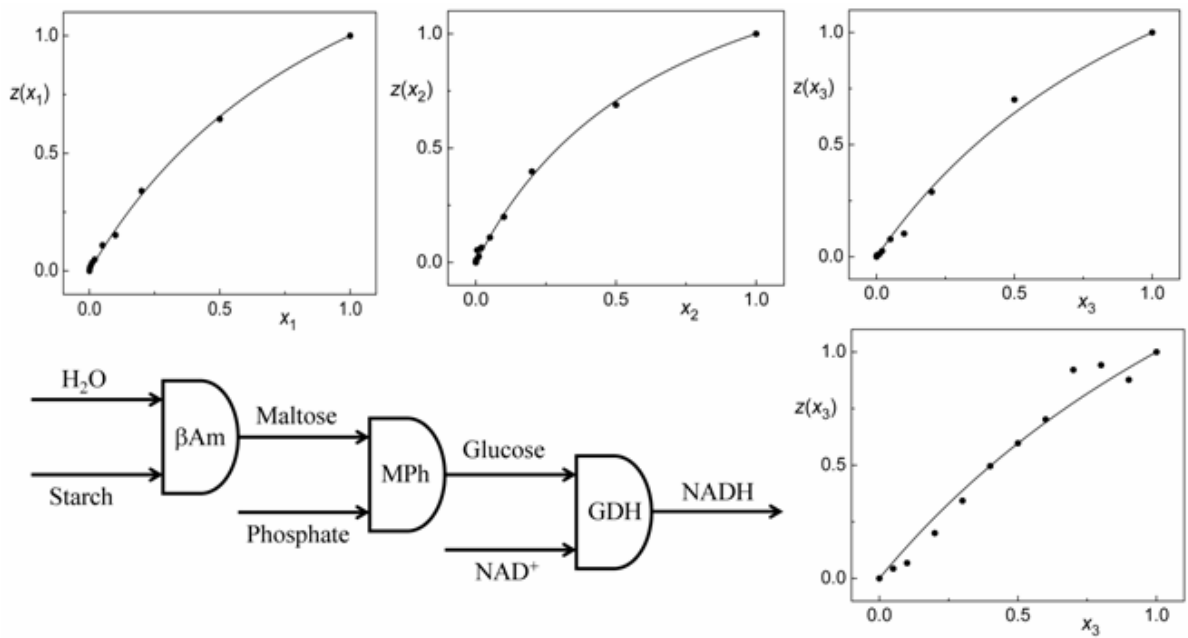

**Fig. 10.** Bottom-left panel: Network of AND gates carried out by β-amylase (βAm), followed by action of maltose phosphorylase (MPh), and then by glucose dehydrogenase (GDH), with the output detected optically. Such cascades are of the type of interest in sensors for dissolved maltose and starch [42]. Top panels: data fits, as detailed in [37], of the measured outputs $z(x_3), z(x_2), z(x_1)$, with all the other inputs held at their logic-1 values. Bottom-right panel: Example of data fit [37] for an experiment with the optimized choice of parameters.

## 5. Electrochemical interfaces for transduction of biochemical signals

We have made significant contributions to the development of biocatalytic and bioaffinity electrochemical sensors for biomedical, environmental and security applications [42,43], aimed at the development of novel nanoparticle-based bioelectronic assays and powerful amplification schemes [44]. Research on biocatalytic electrochemical sensors has aimed at addressing fundamental challenges to practical applications of these devices. These included minimization of the sensor fouling and related biocompatibility issues [45], operation under extreme conditions (e.g., temperature, acidity) by confining the biocatalyst within highly hydrophobic electrode environment [46,47], minimization of interferences between co-existing analytes using tailored permselective coatings [48,49], and attention to oxygen deficiency via new oxygen-rich composite materials [50]. We also developed effective amperometric detection schemes for common outputs expected in enzyme logic-gate based biosensors. This involved the use of metalized-carbon transducers for low-potential highly selective monitoring of hydrogen peroxide [51] and a highly stable transduction of the NADH oxidation at carbon-nanotube modified electrodes [52].

Recent research has illustrated for the first time the electrochemical transduction of enzyme logic gates [25,26]. In particular, we demonstrated an enzyme-logic electronic sensing systems for assessing the overall physiological condition during an injury. The first system consisted of **AND/XOR** logic gates based on the concerted operation of tyrosinase, glucose oxidase, lactate dehydrogenase and microperoxidase-11 [25]. Amperometric detection was demonstrated to be extremely useful for monitoring the formation of the output signals generated by the enzyme logic gates, see Fig. 11. Similarly, electrochemical transduction was applied successfully for another biocomputing system relevant to trauma and shock, based on **AND/IDENTITY** logic gates [26], see Fig. 12. Here, a system composed of three enzymes: lactate oxidase, horseradish peroxidase and glucose dehydrogenase was designed to process biochemical information related to the normal and pathophysiological conditions. The enzymatically processed biochemical information allowed distinguishing between normal physiological conditions and pathophysiological conditions corresponding to traumatic brain injury and hemorrhagic shock. The electrochemical and optical transduction of the enzymatically processed signals was consistent for 8 different combinations of the three input signals, see Fig. 13. While this enzyme-logic bioelectronic research involved immersion of the electrode system into a small-volume sample containing the biocomputing "machinery" along with the corresponding inputs, future efforts should concentrate on understanding of interfacing the biocomputing reagent layers with sensing electrodes for efficient transduction of the output signals in connection with real-life sensing applications. Preliminary efforts have also been devoted to improving microfabrication of electrochemical transducers for monitoring the enzyme-logic outputs [53,54] and towards 'Sense/Act' feedback-loop systems [55]. In particular, flexible thick-film electrodes have been designed recently towards minimally-invasive enzyme-logic bioelectronic sensing [53], see Fig. 14. The influence of the mechanical bending, rolling and crimping of flexible screen-printed electrodes upon their electrochemical behavior has been examined [54].

## 6. Future research directions: biochemical logic systems

Most of the previously studied *chemical* computing systems were controlled by physical input signals (optical, electrical, magnetic, etc.) aiming at molecular computing rather than biosensing applications [1-4]. Introduction of *biochemical* tools for the novel biocomputing systems (particularly enzyme-based logic) opened a new possibility for logic processing of biochemical signals rather than physical ones, thus allowing novel biosensors with integrated biocomputing components. It should be noted that this novel application of biocomputing systems is an emerging development: Only two papers demonstrating biosensors with built-in Boolean logic operations performed by enzymes prior to the electronic transduction of the signals were published to date [25,26]. However, these first examples of biosensors based on the enzyme logic involved a simple logic design, limited optimization of the information processing steps and the absence of physical integration with the transducing interfaces (the enzyme systems were applied in solutions). Further development of novel biosensor concepts based on biocomputing principles should include optimization of the signal processing steps, particularly upon scaling up the system complexity, development of systems specifically relevant for biosensing (using signals meaningful for biomedical, environmental and defense applications) and integration of the biocomputing and signal transduction units into single devices. An additional direction should utilize direct coupling of the biochemical signal processing steps with signal-responsive chemical actuators to yield "smart" systems carrying out micro-mechanical or other functions based on inputs from the biochemical environment, for instance, opening a drug-releasing membrane when a specific pattern of biochemical signals is received.

When the enzyme logic gates were developed earlier for illustrating their computing abilities, chemical signals without specific relevance for their biosensing role were used, selected based on their convenience for the experiments. For the use in digital biosensors, systems should be specifically tailored to accept signals relevant for use in sensor applications. This is expected to impose strict limitations upon the selection of the signal processing enzymes. Most of the chemical/biochemical logic systems designed before operated with **0** logic values represented by the absolute absence of the respective chemicals, while **1** logic values were selected as conveniently high concentrations of chemicals. This was acceptable for the purpose of computing operations, but in the case of the digital biosensors (particularly in biomedical applications) the real background/normal concentrations and analytically meaningful physiological concentrations should be used for **0** and **1** logic values, respectively. This implies restrictions on the definition of the **0**/**1** logic inputs/outputs. In addition, there will be also strict limitations on the level of noise in the signal processing systems, especially for scaling up to multi-component networks.

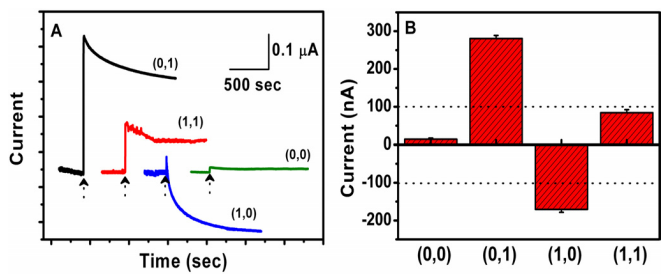

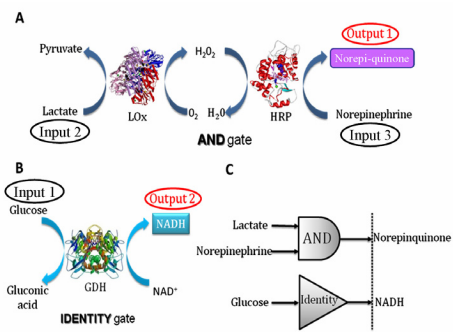

**Fig. 11.** (A) Amperometric responses produced by the **XOR** logic gate generated upon application of different combinations of the input signals: glucose and lactate. (B) Bar diagram featuring the logic **XOR** operation of the system. Adapted from [25].

**Fig. 12.** (A) Multi-input enzyme logic system producing *in situ* pH changes as the output signal. (B) The equivalent logic circuitry for the biocatalytic cascade. Adapted from [26].

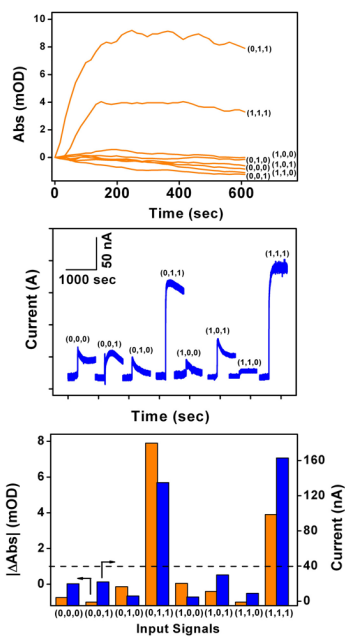

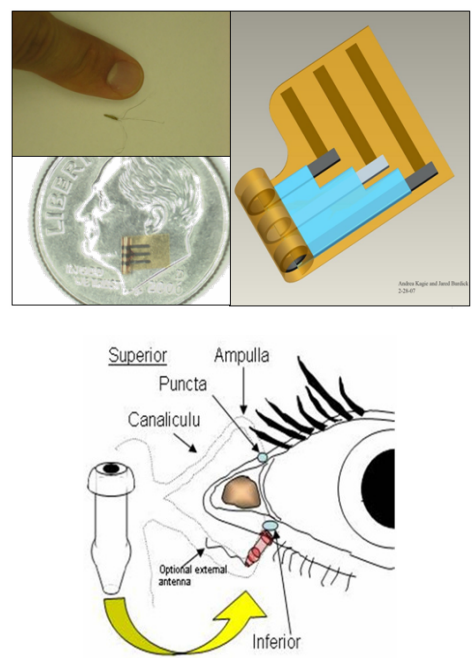

**Fig. 13.** Optical (A) and electro-chemical (B) read out of signals processed by the enzyme logic network (Fig. 12). Digital processing of the input signal combinations. Dash line shows the thresholds separating digital **0** and **1** output signals produced by the system. Adapted from [26].

**Fig. 14.** Microfabricated thick-film, flexible, punctual plug, electrochemical biosensor for insertion into the lacrimal canniculus (tear duct) and monitoring of injury biomarkers in the tear fluid.

It should be noted that the limit of the biocomputing network complexity is set by the cross-reactivity of the enzyme-catalyzed reactions. Only enzymes belonging to different biocatalytic classes (oxidases, dehydrogenases, peroxidises, hydrolases, etc.) could operate in a single "soup" without significant cross-reactivity. If chemical reasons require the use of cross-reacting enzymes in the system, they must be compartmentalized using surface patterning or in microfluidic devices. Application of more selective biomolecular interactions would offer the advantage of making biocomputing systems more specific to various input signals and less cross-reactive in the chemical signal processing.

## 7. Future research directions: modeling of biochemical logic networks

Presently available research results on gate and network modeling, analysis and optimization for enzyme-based and similar biochemical computing systems, briefly surveyed in Sect. 4, have been limited to detailed rate-equation analysis of single **AND** gates, and to small network optimization be changing the relative activity of gates in it [36-41]. Future research should explore a comprehensive approach for optimization of networks for biosensing, relying on earlier results but incorporating new components for analog/digital error suppression for larger networks. Specifically, for multi-input systems, we favor the following design strategy.

*Step 1: Modular network analysis.* The enzymatic cascade will be represented as a Boolean network. Networked gates will then be parameterized in terms of response functions typical for the specific gate or reaction. Presently, we already have experience [37] with networked **AND** gates, with the response surface of each gate fitted to the form $F(x,y) = xy(1+a)(1+b)/(x+a)(y+b)$ with two adjustable parameters, $0 < a, b \leq \infty$ (see [37] for details).

Similar fitting forms can be devised, guided by the properties of the biochemical reactions involved, for Boolean-gate (e.g., **OR**) and other network elements, such as filters, signal splitters and balancers, feedback loops, etc.

*Step 2: Detailed network representation and adjustment of relative component activities.* At this stage, the network functioning will be analyzed for analog noise amplification and propagation. This will include adjustment of the gate activities to diminish the effects of the "noise amplifying" ones [37,41], but also a more careful network structure analysis in terms of not only Boolean, but other network elements.

*Step 3: Gate function optimization.* The key gates in the network, will be analyzed in detail for optimization based on the kinetic, rate-equation description. Gate optimization involves the use of few-parameter (in some cases even simpler than the standard Michaelis-Menten kinetics) description of the enzymatic-reaction kinetics, to parameterize the gate response function, $F(x,y)$, the slopes (and generally the shape) of which near the logic points **00**, **01**, **10**, **11** determine the degree of analog noise amplification (or suppression). Details of this approach have been presented in [36,37,39,40]. In complex enzymatic cascades we expect the enzyme(s), of concentration(s) to be collectively denoted $E$, to constitute most of the gate "machinery". Even if we fit in detail the rate constants, say $k_\alpha, k_\beta, ...$, for each gate, then the set of the controllable parameters is $t_{\max}; E; k_\alpha, k_\beta, ...; I_\gamma, I_\delta, ...$. Here $I_\gamma, I_\delta, ...$ stand for input concentrations of possible additional chemicals (which are not used as our "logic" inputs fixed by the environment) taken in by the enzymes at various stages (and possibly at various times, not just at time 0) of the logic-network functioning. The rate constants $k_\alpha, k_\beta, ...$ can be varied primarily by changing the physical or chemical conditions, which might not always be possible in specific applications, especially biomedical ones. Therefore, we can only adjust the set $E; I_\gamma, I_\delta, ...$. However, the dependence on some concentrations, especially of the enzyme itself, $E$, may to a large extent, cancel out in the "logic" dimensionless output, $F$, because the overall gate activity is approximately linear for the enzyme concentration in many regimes of applications (with the enzymatic reactions held near steady state conditions). Thus, we are left with few or no gate parameters for a straightforward gate-by-gate optimization.

*Step 4: The role of non-Boolean network elements.* These considerations point to the conclusion that additional network elements, specifically, filters to generate sigmoidal response, will generally be needed not only for larger network design for digital error correction but even for individual gate optimization. Thus, for complex networks the role of specially designed filter-response steps, and possibly of other elements will become more important in achieving the overall network stability and low-noise functioning.

## 8. Future research directions: integration of signal processing biocomputing systems with electrochemical transducers

Unlike common biosensing devices based on a single input (analyte), devices based on biochemical logic systems require a fundamentally new approach for the sensor design and operation. Future efforts will focus on the interface of biocomputing systems and electrochemical transducers, with particular attention given to the composition, preparation and immobilization of the biocomputing surface layer (see below), to the role of the system scalability and to the efficient transduction of the output signals.

*Surface Immobilization of the Biocomputing Machinery.* As common in conventional electrochemical biosensors, the success of the enzyme-logic biosensor will depend, in part, on the immobilization of the biocomputing reagent layer. Unlike early studies [25,26] where the multiple gate constituents and inputs were dissolved in a solution, biochemical logic sensors require optimal surface confinement of the biocomputing layer. A careful engineering of the enzyme microenvironment (on the surface) is essential for an optimal performance, considering the increased complexity of logic-gate biosensors (compared to common simpler enzyme electrodes). The objective will be not only to provide a contact between the biocomputing layer and the transducing (electrode) surface but also to combine efficiently the individual logic-gate elements (to ensure efficient coupling of the enzyme cascade). Such efficient couplings should be accomplished while maintaining high enzymatic stability and retaining the individual reagents (e.g., avoid leakage of co-substrates). Particular attention will be given to the composition of the "reagent layer," especially to the level of the multiple enzyme and co-substrate components of the logic network machinery, as well as of the corresponding buffer salts and enzyme stabilizers. The selection of the enzymes should also ensure the absence of cross-reactions among the individual biocatalytic gates. Unlike solution experiments where the entire biochemical machinery is homogenously mixed [25,26], the surface confinement may actually allow an ordered (optimal) placement of the individual gates (through a layer-by-layer configuration), hence ensuring a more efficient and rational coupling of the enzyme cascade and avoiding cross-talk. Depending on the specific sensing goal, the level of the surface-confined reagents (enzymes, co-substrates, mediators, etc.) can be tailored to account for potentially largely different input concentrations or for substantially different enzyme activities. The permeability of the coating can also be optimized for tailoring the transport of the individual inputs (and hence tuning their levels), while excluding potential interferences and protecting the surface. Additional protective layers may be required, particularly in connection to biomedical applications.

*Optimal Transduction of Biocomputing Signal Processes.* Electrochemical transduction of the system outputs should also receive careful attention in view of the "digital" character of the system and the possible multiple output signals. Special attention should be given to the electrochemical detection mode, to the selection of the electrode materials, and attainment of optimal signal-to-noise characteristics. Simultaneous measurements of multiple output signals will require new transduction strategies (compared to common single output sensors). Desirably, these could be accomplished using a single working electrode, hence simplifying the design of enzyme-logic biosensors (compared to multi-electrode systems). This may involve a potential scan for simultaneous measurement of the multiple outputs. Alternately, multi-potential steps (pulses) to different values can be optimized for detecting the corresponding outputs.

Attention should also be given to establishing the digital "background" (**0**) values corresponding to "normal" (physiological or environmental) levels of the target analytes. Signal amplification schemes, involving nanoscale materials [44] or product accumulation and recycling [56], can be evaluated for enhancing the sensitivity in connection to ultralow levels of certain inputs (e.g., biomarkers or bioagents). Such amplification schemes will benefit also biocomputing sensors involving a narrow **0** to **1** range.

We acknowledge funding by the National Science Foundation (grants CCF-0726698, DMR-0706209), Office of Naval Research (grant N00014-08-1-1202), and Semiconductor Research Corporation (award 2008-RJ-1839G).

Posted as e-print 0909.1583 at www.arxiv.org

Click this text for updates and for better-resolution images